\documentclass[aps,prb,twocolumn,groupedaddress,showpacs,amsmath,floatfix]{revtex4}
\usepackage{graphicx}
\usepackage{amssymb}

\begin{document}


\title{Spectroscopy of low-frequency noise and its temperature dependence in a superconducting qubit}

\author{Fei Yan$^1$} \email[fyan@mit.edu]{}
\author{Jonas Bylander$^2$}
\author{Simon Gustavsson$^{2}$}
\author{Fumiki Yoshihara$^{3}$}
\author{Khalil Harrabi$^{3}$}
\altaffiliation{Present address: Physics Dept., King Fahd University of Petroleum \& Minerals, Dhahran 31261, Saudi Arabia}
\author{David G. Cory$^{4,5}$}
\author{Terry P. Orlando$^{2,6}$}
\author{Yasunobu Nakamura$^{3,7}$}
\altaffiliation{Present address: Research Center for Advanced Science and Technology (RCAST), University of Tokyo, Komaba, Meguro-ku, Tokyo 153-8904, Japan}
\author{Jaw-Shen Tsai$^{3,7}$}
\author{William D. Oliver$^{2,8}$}

\affiliation{$^1$Department of Nuclear Science and Engineering, Massachusetts Institute of Technology (MIT), Cambridge, Massachusetts 02139, USA
$^2$Research Laboratory of Electronics, MIT, Cambridge, Massachusetts 02139, USA
$^3$Institute of Physical and Chemical Research (RIKEN), Wako, Saitama 351-0198, Japan
$^4$Institute for Quantum Computing and Department of Chemistry, University of Waterloo, ON, N2L 3G1, Canada
$^5$Perimeter Institute for Theoretical Physics, Waterloo, ON, N2J, 2W9, Canada
$^6$Department of Electrical Engineering and Computer Science, MIT, Cambridge, Massachusetts 02139, USA
$^7$Green Innovation Research Laboratories, NEC Corporation, Tsukuba, Ibaraki 305-8501, Japan
$^8$MIT Lincoln Laboratory, 244 Wood Street, Lexington, Massachusetts 02420, USA}


\date{\today}

\begin{abstract}
We report a direct measurement of the low-frequency noise spectrum in a superconducting flux qubit.
Our method uses the noise sensitivity of a free-induction Ramsey interference experiment, comprising free evolution in the presence of noise for a fixed period of time followed by single-shot qubit-state measurement.
Repeating this procedure enables Fourier-transform noise spectroscopy with access to frequencies up to the achievable repetition rate, a regime relevant to dephasing in ensemble-averaged time-domain measurements such as Ramsey interferometry.
Rotating the qubit's quantization axis allows us to measure two types of noise: effective flux noise and effective critical-current or charge noise.
For both noise sources, we observe that the very same $1/f$-type power laws measured at considerably higher frequencies ($0.2-20$\,MHz) are consistent with the noise in the $0.01-100$-Hz range measured here.
We find no evidence of temperature dependence of the noises over $65-200$\,mK, and also no evidence of time-domain correlations between the two noises.
These methods and results are pertinent to the dephasing of all superconducting qubits.

\end{abstract}

\pacs{03.67.Lx, 74.40.-n, 74.25.Sv, 85.25.Cp, 85.25.Dq}


\maketitle

\section{Introduction}

A major remaining obstacle to implementing fault-tolerant quantum computation with superconducting qubits is the insufficient coherence time $T_2$ compared to the gate-operation time.
%
In the Bloch--Redfield picture of two-level system dynamics,
there are two mechanisms that limit $T_2$:
energy relaxation $T_1$ due to noise at the transition frequency $\nu_{01}$,
and dephasing $T_\varphi$ due to low-frequency fluctuations of $\nu_{01}$.
In the qubit's energy eigenbasis, these contributions are due to the transverse and longitudinal components of the fluctuations, respectively.
In cases when both relaxation and dephasing exhibit exponential decay laws, their inverse times add to a decoherence rate
$T_2^{-1} = (2 T_1)^{-1} + T_\varphi^{-1}$.
While $T_1$ can now exceed $10\,\mu$s in several superconducting qubit modalities~\cite{Kim-PRL-11,Bylander-NPhys-2011,Paik-PRL-11}, further improvements are required to comfortably exceed even the most lenient error correction thresholds.
In general, energy relaxation is irreversible, such that further control-based improvements require resource-intensive multi-qubit quantum error-correction protocols.
Dephasing, on the other hand, can be refocused by dynamical-decoupling techniques~\cite{Bylander-NPhys-2011}, with only a modest amount of resource overhead.
The ultimate goal is to mitigate and eliminate the noise leading to both types of decoherence.
To this end, a more detailed understanding of the noise processes -- such as magnetic-flux, critical-current, and charge fluctuations -- would expedite materials science, device engineering, and the development of coherent-control methods.

Effective surface spins have recently been identified as one dominant source of low-frequency magnetic-flux noise~\cite{Sendelbach-PRL-08,Sendelbach-PRL-09}, detrimental to several types of superconducting qubits;
however, open questions remain regarding the nature of these spins.
Their noise is known to be due to local fluctuators~\cite{Koch-PRL-07,Choi-PRL-09,Yoshihara-PRB-10,Gustavsson-11} and the spectrum exhibits a $1/f^\alpha$ power-law dependence from hertz to tens of megahertz~\cite{Wellstood-APL-87,Ithier-PRB-05,Yoshihara-PRL-06,Bialczak-PRL-07,Faoro-PRL-08}.
Its dependence on the device geometry~\cite{Lanting-PRB-09,Wellstood-IEEE-11} merits further study.

Similarly, for the flux qubit~\cite{Orlando-PRB-99,Mooij-Science-99}, the noise in the tunnel coupling $\Delta$ between the persistent-current states shows a $1/f$-type spectrum from hertz to hundreds of kilohertz~\cite{Bylander-NPhys-2011}.
This noise may originate in the critical-current fluctuations of the Josephson junctions~\cite{vanHarlingen-PRB-04,Wellstood-APL-04,Muck-APL-05,Eroms-APL-06},
and/or fluctuating offset charges, due to, e.g., charge traps located in the oxides of the junction, metal--insulator interfaces, or surfaces.
Charge noise can lead to dephasing even in the flux qubit, even though the junctions have a relatively high ratio of Josephson-tunneling to Coulomb-charging energies ($E_\mathrm{J}/E_\mathrm{C} \approx 50$ in our device).

In this paper, we introduce a measurement technique for low-frequency noise.
%
A distinguishing feature of our technique is that it enables the measurement of noise spectra up to frequencies limited only by the achievable measurement repetition rate.
This is important, because noise measured in this manner resides (at least in part) within the relevant measurement bandwidth of time-domain experiments, e.g., Ramsey interferometry, that use the standard ensemble-averaging (that is, the averaging of multiple trials acquired at the same repetition rate) to estimate the qubit-state occupation probability.

We report on a direct characterization of the $1/f$-noise-power spectral densities (PSD) $S(f)$ in an aluminum superconducting flux qubit~\footnote{The Al/AlO$_x$/Al device was made by shadow evaporation at NEC; the experiments were performed at MIT.}.
We distinguish between the two noises $\delta\varepsilon$, which is effective flux noise, and $\delta\Delta$, which can be parameterized as effective critical-current noise or effective charge noise.
Interestingly, we find that the same $1/f^\alpha$ power laws, measured at much higher frequencies~\cite{Bylander-NPhys-2011}, extend down to the $10^{-2}-10^{+2}$-Hz range nearly unchanged.
Over the temperature range
$65-200$\,mK, both noises, $\delta\varepsilon$ and $\delta\Delta$, are  independent of temperature,
and any $\delta\varepsilon-\delta\Delta$-noise correlations are very small or non-existent.


\section{Experimental methods, analysis, and results}

\begin{figure}      
\centering
\includegraphics[width=7.5cm]{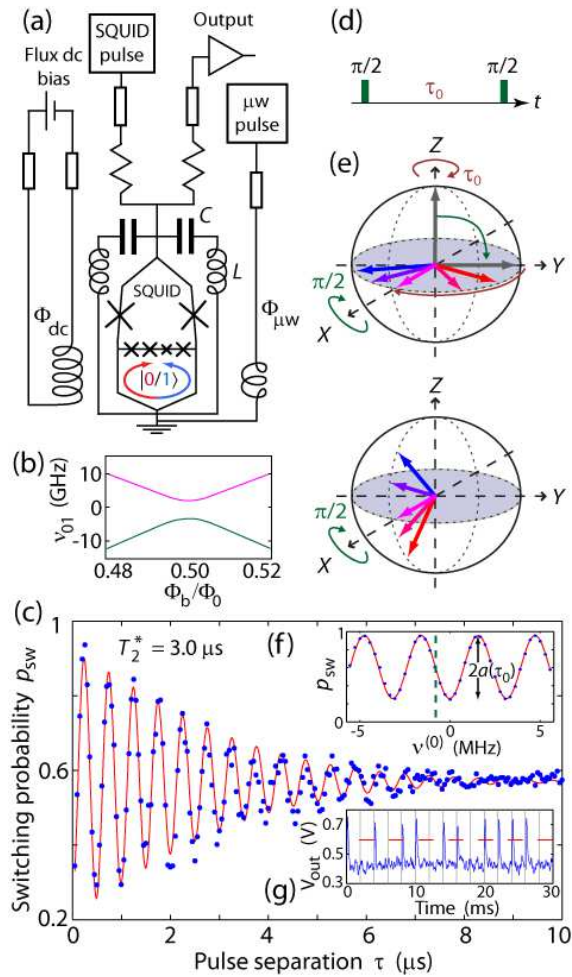}
  \caption{\label{fig:deviceDescription} (color online)
  \textbf{(a)} Schematic of the qubit device with biasing and read-out circuitry.
  %
  \textbf{(b)} Qubit energy-level diagram near
  $\Phi_\mathrm{b} = \Phi_0/2$ ($\varepsilon=0$).
  \textbf{(c)} Ramsey interference fringe measured at $\varepsilon = 0$ with  $5,000$ averages per point.
  %
  \textbf{(d)} Double-$\pi/2$ pulse (Ramsey) sequence.
  \textbf{(e)} Bloch sphere representation of dephasing and our measurement scheme:
  the initial $\pi/2$ pulse takes the Bloch vector from the ground state (along $Z$) to the equator (along $Y$);
  during the free-induction time $\tau=\tau_0$, the nominal detuning is $\nu^{(0)} = 1/4\tau_0$, but differs slightly at each trial (quasi-static noise), and the Bloch vectors therefore acquire different phases;
  the final $\pi/2$ pulse rotates the Bloch vectors out of the equatorial plane so that they can be distinguished by the SQUID read out.
  \textbf{(f)} Calibration by scanning $\nu_{\mu\mathrm{w}}$ with $\tau_0=0.3\,\mu$s, $\varepsilon=0$.
  The red, solid line is a sinusoidal fit used to determine the working point for maximal sensitivity (dashed line at $\nu^{(0)}=-0.8$\,MHz).
  \textbf{(g)} Individual SQUID-switching events measured on an oscilloscope. The dashed, horizontal line is a software threshold detector.
}
\end{figure}

The flux qubit's two-level Hamiltonian is
$\hat{\mathcal{H}} = - (h/2) \left(\varepsilon\hat{\sigma}_{x} + \Delta\hat{\sigma}_{z} \right)$.
Here $\varepsilon = 2I_\mathrm{p}\Phi_\mathrm{b}/h$ is the energy detuning between the diabatic states of classical circulating current $I_\mathrm{p}=0.18\,\mu$A, and $\varepsilon$ is adjusted by the external magnetic flux $\Phi$ via
$\Phi_\mathrm{b} = \Phi-\Phi_0/2$ ($\Phi_0=h/2e$ is the superconducting flux quantum);
see the schematic in Fig.~\ref{fig:deviceDescription}(a--b).
$\Delta=5.4$\,GHz is the tunnel coupling that hybridizes the persistent-current states, and is established during fabrication by the $E_\mathrm{J}/E_\mathrm{C}$ ratio.
We write each of the parameters $\lambda=\varepsilon,\,\Delta$ as the sum of its nominal value and a time-dependent fluctuation, $\lambda(t) = \lambda^{(0)} + \delta\lambda(t)$.
We distinguish between the effects of $\delta\varepsilon$ and $\delta\Delta$ fluctuations by rotating the qubit's quantization axis (eigenbasis),
thereby altering the sensitivity of the energy-level splitting
$\nu_{01} = \sqrt{\varepsilon^2 + \Delta^2}$
to fluctuations,
$D_\lambda = \partial\nu_{01}/\partial\lambda = \lambda/\nu_{01}$; see Fig.~\ref{fig:deviceDescription}(b).
The dominant contributor to longitudinal fluctuations in the qubit's energy eigenbasis is $\delta\Delta$ noise at $\varepsilon=0$ (as the second-order contribution from $\delta\varepsilon$ noise is negligible~\cite{Bylander-NPhys-2011}), and $\delta\varepsilon$- (flux) noise for $|\varepsilon|\gtrsim 0.1$\,GHz.
%


We use a hysteretic SQUID to read out the qubit's state.
During qubit manipulation, we conveniently set the SQUID current $I_\mathrm{b}$ to zero. (This is close to its optimal value of $I_\mathrm{b}^* = 50$\,nA in this device, at which $\delta I_\mathrm{b}$ fluctuations are decoupled from the qubit~\cite{Bertet-PRL-05}.)


\begin{figure*}      
\centering
\includegraphics[width=15cm]{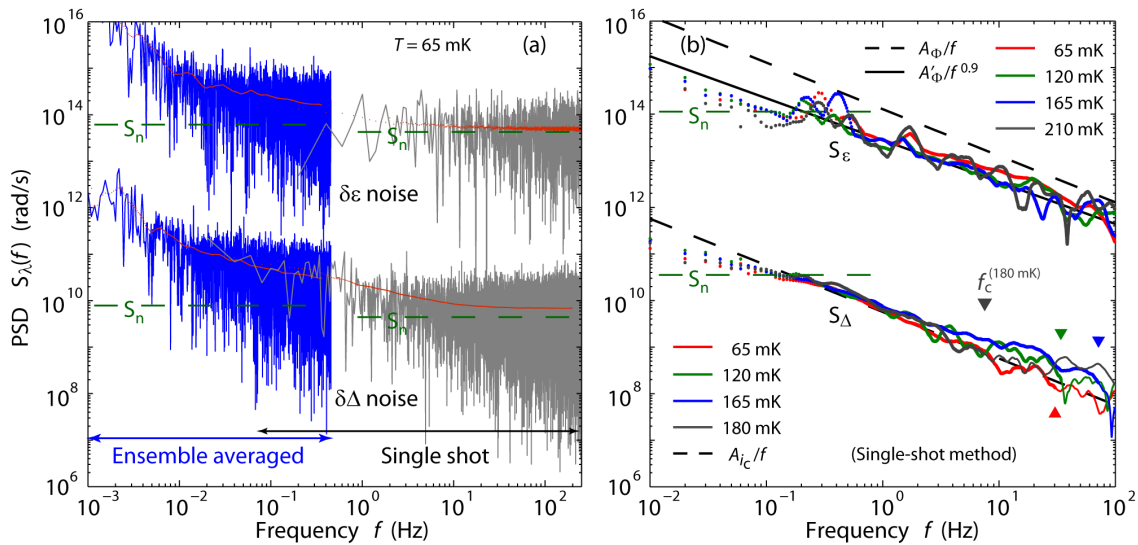}
  \caption{\label{fig:EpsDeltaNoise_vs_Temp} (color online)
  Bilateral noise PSDs.
  (a) Standard-method PSD of the ensemble-averaged time series of switching probabilities $p_\mathrm{sw}$
  (blue data, on left) and of the binary time series of single-shot measurements $\{z_n\}$ (gray, on right), measured at base temperature.
  The red points (thin line) are averages of hundreds of traces.
  ($S_\varepsilon$ measured at $\varepsilon=450$\,MHz.)
  (b) Cross-PSDs of interleaved time series, $S_\lambda(f)$ (Eq.~\ref{eq:psd_fqUnits}), vs.\@ device temperature ($S_\varepsilon$ measured at $\varepsilon=640$\,MHz).
  The data were smoothed by a sliding average with a triangular weight function of width $\Delta f = f/4$.
  The dashed, diagonal lines are the $1/f$ noises
  $S_\lambda (f) = (2\pi)^2 \kappa_\lambda^2 A_{\Phi,i_\mathrm{c}}/|f|$,
  derived in Ref.~\onlinecite{Bylander-NPhys-2011};
  the sensitivities are
  $\kappa_{\Delta,i_\mathrm{c}} \equiv \partial\Delta/\partial i_\mathrm{c} = 3.2\,\mathrm{GHz}$ ($i_\mathrm{c}$ is normalized critical current)
  and
  $\kappa_\varepsilon \equiv \partial\varepsilon/\partial\Phi = 1.1\,\mathrm{GHz/m}\Phi_0$,
  and the noise strengths are
  $A_{i_\mathrm{c}} = (4.0\times10^{-6})^2$
  and
  $A_\Phi = (1.7\,\mu\Phi_0)^2$.
  The solid, diagonal line is $S_\varepsilon(f) = (2\pi)^2 \kappa_\varepsilon^2 A'_\Phi/|f|^{0.9}$,
  with $A'_\Phi = (0.8\,\mu\Phi_0)^2$.
  See Ref.~\onlinecite{Bylander-NPhys-2011} and its supplementary information for details.
  The horizontal, green, dashed lines are the sampling-noise levels $S_\mathrm{n}(\tau_0,T)$ at low temperature; the triangles represent an upper cut-off frequency $f_\mathrm{c}$ for sufficient averaging, above which the data are not dependable.
  %
  }
\end{figure*}

To directly probe the fluctuations, we repeatedly let the qubit undergo Ramsey free induction; see Fig.~\ref{fig:deviceDescription}(c).
Instead of scanning the pulse separation $\tau$, we fix it at a value $\tau_0$, chosen to maximize the qubit state's sensitivity to noise.
We also fix the nominal detuning $\nu^{(0)}$ of the applied microwave frequency from the qubit's frequency to $\nu^{(0)}\equiv\nu_{01}-\nu_{\mu\mathrm{w}} = 1/4\tau_0$ (a free-evolution $\pi/2$ rotation in the $X\!\!-\!\!Y$ plane); see Fig.~\ref{fig:deviceDescription}(d--f).
The $\delta\lambda(t)$ fluctuations translate into frequency fluctuations, $\nu(t) = \nu^{(0)} + \delta\nu(t)$, which in turn translate into fluctuations of the SQUID's switching probability,
$p_\mathrm{sw}(\nu,\tau_0) = p_0 - a_0(T)\,\cos(2\pi\,\nu(t)\,\tau_0)$,
where $a_0(T)$ is the temperature-dependent read-out visibility ($2\,a_0=79$\,\% at the refrigerator's $T=12$-mK base temperature) and $p_0$ is the switching probability for the qubit's 50-\% superposition state.
We linearize about a working point at $p_\mathrm{sw} = p_0$ and obtain
$\partial p_\mathrm{sw}/\partial\nu = 2\pi \, a_0(T) \,\tau_0$.
Due to this transfer function, a correction factor arises in the calculation of ensemble-averaged quantities, assuming Gaussian statistics
(see section B.2 of the appendix).
This factor is $a(\tau_0,T)/a_0(T)$, where $a(\tau_0,T)$ is the amplitude of the fringe at pulse separation $\tau_0$.
%
The conversion factor from the noise $\delta \lambda$ to the switching probability $p_\mathrm{sw}$ then becomes
$\eta_\lambda(\tau_0,T) = 2\pi \, a(\tau_0,T) \,\tau_0 D_\lambda$.


The single-shot read-out of the qubit, with a repetition time $\Delta t$, results in a binary time series $\{z_n\}$; see Fig.~\ref{fig:deviceDescription}(g).
(In this experiment, $\Delta t=2$\,ms to allow for the read-out induced quasiparticles to relax between trials so that they contribute negligibly to heating.)
Each element $z_n$ represents the result of a Bernoulli trial with expectation value $p_\mathrm{sw}$.
%
%
The standard method to determine the noise-PSD \cite{Bialczak-PRL-07,Lanting-PRB-09,Sank-arxiv-11}
is to ensemble average the switching events acquired during a gate time
$t_\mathrm{acq} = N_\mathrm{G}\Delta t$,
with typically $N_\mathrm{G}\approx 1,000$,
to determine the average switching probability $p_\mathrm{sw}$ (binomially distributed),
and then take the Fourier transform of the time series of switching probabilities $p_\mathrm{sw}$; see Fig.~\ref{fig:EpsDeltaNoise_vs_Temp}(a).
%
This approach estimates the noise spectrum at frequencies up to $1/2t_\mathrm{acq}$.
In contrast, in our approach, we calculate the bilateral PSD from the recorded series of single-shot measurements (Bernoulli trials),
$S(f_k) = |Z_k|^2 / (N/\Delta t)$, where $f_k = k/N\Delta t$,
$k = 0,\ldots, N/2$, and $\{Z_k\}$ is the discrete Fourier transform~\footnote{Contrary to Refs.~\onlinecite{Ithier-PRB-05,Yoshihara-PRL-06,Bylander-NPhys-2011}, we use the type-1 Fourier transform, $S_x(f) = \int_{-\infty}^\infty \mathrm{d}t \langle x(0)x(t) \rangle \exp(-i2\pi ft)$.}
of $\{z_n\}$,
with $n$ typically ranging from 1 to $N=5\times 10^4$.
This method increases the upper cut-off frequency from $1/2N_\mathrm{G}\Delta t$ to $1/2\Delta t$, which may approach $1/\tau_0$ and is limited only by the achievable repetition rate; see Fig.~\ref{fig:EpsDeltaNoise_vs_Temp}(a--b), section A of the appendix, and Fig.~\ref{fig:PSD_Sketch}.

Both the PSDs originating from single-shot and from ensemble-averaged measurements in Fig.~\ref{fig:EpsDeltaNoise_vs_Temp}(a) exhibit statistical-sampling noise
$S_\mathrm{n}(\tau_0,T) = (2\pi)^2 \,\sigma_\mathrm{s}^2\,\Delta t/\eta_\lambda^2(\tau_0,T)$, where the variance is $\sigma_\mathrm{s}^2 = p_\mathrm{sw}(1-p_\mathrm{sw})$.
We can eliminate this white background noise by calculating the bilateral cross-PSD of the two interleaved (single-shot) time series $z'_n = z_{2n-1}$ and $z''_n = z_{2n}$,
\begin{equation} \label{eq:cross_psd_2}
  S^\mathrm{cross}(f_{k}) = \frac{Z'_k (Z''_k)^*}{N/\Delta t} , \quad
  f_k = \frac{k}{N\Delta t} ,
\end{equation}
where now $k = 0,\ldots, N/4$.
Dividing Eq.~(\ref{eq:cross_psd_2}) by the conversion factor, we obtain the spectral density of the fluctuation $\lambda$,
\begin{equation} \label{eq:psd_fqUnits}
  S_\lambda(f_k) = (2\pi)^2\,S^\mathrm{cross}(f_k)/\eta_\lambda^2(\tau_0,T) .
\end{equation}
We typically average the spectra of 500 time series to improve statistical accuracy without compromising bandwidth, and recalibrate the working point periodically (hourly).
Note that both the $1/f$ noise and the sampling noise dominate all other background noise at the temperatures considered.
%


The $\delta\varepsilon$ and $\delta\Delta$ noise-PSDs are plotted in Fig.~\ref{fig:EpsDeltaNoise_vs_Temp}(b) for several temperatures.
There is striking agreement with the $1/f^\alpha$ power laws inferred in Ref.~\onlinecite{Bylander-NPhys-2011}, measured at considerably higher frequencies ($0.02-20$\,MHz).
%
%
Noise that is strictly $1/f^{\alpha=1}$ over the frequency range relevant to free-induction (here $10^{-1} \!\sim \!10^6$\,Hz) gives a Gaussian decay function of the temporal Ramsey oscillations.
Assuming that our $\delta\Delta$ noise satisfies these criteria, we use the approach of Ref.~\onlinecite{Martinis-PRB-03} to calculate the inhomogeneously broadened decay-time constant $T_\varphi^*$.
With noise sensitivity $\kappa_{\Delta,i_\mathrm{c}}$ and strength $A_{i_\mathrm{c}}$ as defined in the legend of Fig.~\ref{fig:EpsDeltaNoise_vs_Temp}, $T_1=12\,\mu$s, and typical parameters $\tau_0=1\,\mu$s and $t_\mathrm{acq}=10$\,s, we obtain
$T_\varphi^* = \left(2\pi\,\kappa_{\Delta,i_\mathrm{c}} D_\Delta\right)^{-1} A_{i_\mathrm{c}}^{-1/2} \left(\ln[t_\mathrm{acq}/2\,\tau_0]\right)^{-1/2} = 3.2\,\mu$s, in very good agreement with the observed $T_\varphi^*$ in Fig.~\ref{fig:deviceDescription}(c).

\begin{figure}      
\centering
\includegraphics[width=8cm]{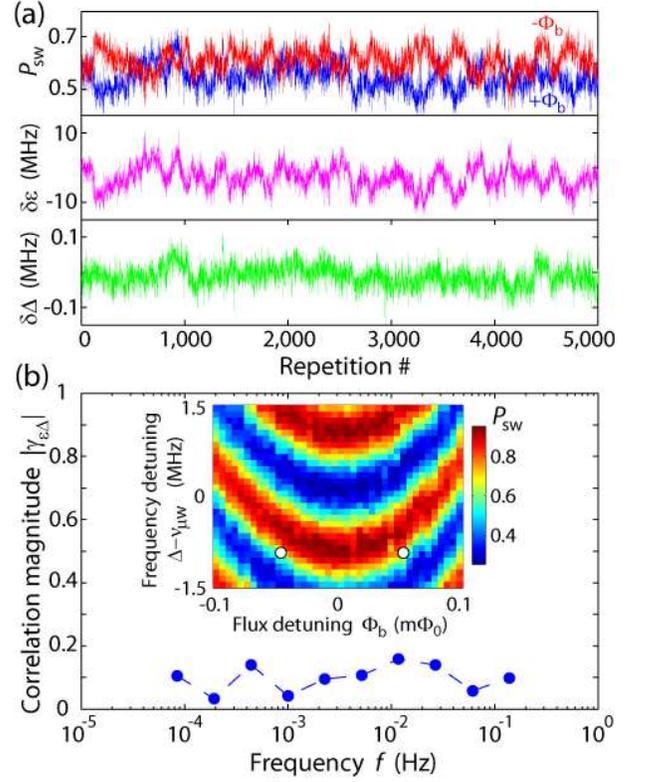}
  \caption{\label{fig:correlations} (color online)
  $\delta\varepsilon-\delta\Delta$-noise correlations measured at base temperature.
  \textbf{(a)} \textbf{Top panel:} Time series of $p_\mathrm{sw}$ with fixed $\tau_0=0.3\,\mu$s, nominal $\nu^{(0)}=-0.8$\,MHz, and 1,000 averages per measured point.
  The flux-bias polarity ($\pm\varepsilon^{(0)}$) was alternated between positive (blue) and negative (red) [$\pm\Phi_\mathrm{b}=\pm0.058\,\mathrm{m}\Phi_0$,
  see inset in (b)],
  with a 2-s repetition period.
  \textbf{(a)} \textbf{Middle and bottom panels:} Differential ($\delta\varepsilon$) and common-mode ($\delta\Delta$) noise inferred from the data in the top panel.
  \textbf{(b)} Magnitude of the correlation function, (Eq.~\ref{eq:corr_ampl});
  the smaller the magnitude, the more indeterminate the phase, and therefore we cannot discriminate between possible correlation and anticorrelation.
  \textbf{Inset:} 30-min measurement of Ramsey fringes
  (frequency detuning referred to $\varepsilon=0$, where $\nu_{01} = \Delta - \nu_{\mu\mathrm{w}}$).
  White circles indicate the bias points used in the top panel of (a).
}
\end{figure}

We now turn to possible $\delta\varepsilon-\delta\Delta$-noise correlations in the time domain, to check that the two spectra in Fig.~\ref{fig:EpsDeltaNoise_vs_Temp} are not due to one and the same mechanism.
Figure~\ref{fig:correlations} shows how we repeatedly measured the switching probability at alternating flux biases,
$p_\mathrm{sw}(\pm\varepsilon^{(0)}+\delta\varepsilon, \Delta^{(0)}+\delta\Delta)$,
with $\pm\varepsilon^{(0)}$ chosen such that the effects of the two noises on $\delta\nu$ were similar in magnitude,
i.e.\@ $D_\varepsilon\left(\varepsilon^{(0)}\right)|\delta\varepsilon| \approx D_\Delta\left(\varepsilon^{(0)}\right)|\delta\Delta|$.
We set the pulse separation $\tau_0$ and nominal frequency detuning $\nu^{(0)}$.
With the
energy-level splitting $\nu(\varepsilon,\Delta) = \nu^{(0)}\left(\varepsilon^{(0)},\Delta^{(0)}\right) + \delta\nu\left(\varepsilon^{(0)},\delta\varepsilon,\delta\Delta\right)$,
we use the decay function of the Ramsey fringe to infer the noise correlations from the measured $p_\mathrm{sw}$,
\begin{equation} \label{eq:Ramsey_decay_fcn}
 p_\mathrm{sw}(\varepsilon, \Delta) = p_0 - a_0 \,
 \exp\left( -\tau_0/2T_1 \right) \times
\end{equation}
$$
 \qquad \times \exp\left( -\left[\tau_0/T_\varphi^*(\varepsilon)\right]^2 \right)
 \cos\Big( 2\pi\,\nu(\varepsilon,\Delta) \,\tau_0 \Big) ,
$$
where
$[1/T_\varphi^*(\varepsilon)]^2 = [1/T_\varphi^*(0)]^2 + K A_\Phi D_\varepsilon^2$ and $K$ is a constant that we have determined independently, along with the other parameters in the equation.
At the bias points $\varepsilon=\pm\varepsilon^{(0)}$, $\delta\varepsilon$ fluctuations induce negatively correlated $\delta\nu$ fluctuations (and consequently $p_\mathrm{sw}$ fluctuations), whereas $\delta\Delta$ fluctuations induce positively correlated $\delta\nu$ fluctuations.
At each time step, the measurement of $p_\mathrm{sw}$ for $\varepsilon=\pm\varepsilon_0$ yields a system of two non-linear equations in the two unknowns $\delta\varepsilon$ and $\delta\Delta$.
We solve this system numerically: Fig.~\ref{fig:correlations}(a) shows the raw $p_\mathrm{sw}$ data and extracted $\delta\varepsilon$ and $\delta\Delta$ vs.\@ time.
We then calculate the cross-PSD $S_{\varepsilon\Delta}(f)$ as the Fourier transform of the cross-correlation function, and obtain an upper bound on the magnitude of the correlation function, as shown in Fig.~\ref{fig:correlations}(b),
\begin{equation} \label{eq:corr_ampl}
|\gamma_{\varepsilon\Delta}(f)| =
\frac{|S_{\varepsilon\Delta}(f)|}{[S_{\varepsilon}(f)S_{\Delta}(f)]^{1/2}} < 0.2 .
\end{equation}
%


\begin{figure}      
\centering
\includegraphics[width=8cm]{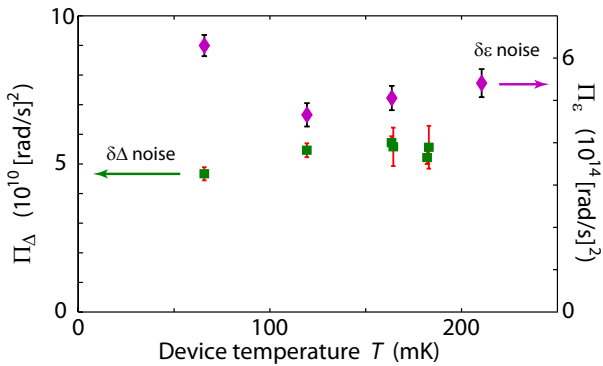}
\vspace{0mm}
  \caption{\label{fig:tot_noise_power} (color online)
  Temperature dependencies of the total noise powers
  $\Pi_\lambda(T) = (2\pi)^2 \kappa_\lambda^2 \int_{F_\lambda} \mathrm{d}f \, S_\lambda(f; T)$
  in the frequency intervals
  $F_\varepsilon = 0.02 - 50$\,Hz and $F_\Delta = 0.02 - 2$\,Hz, cf.\@ Fig.~\ref{fig:EpsDeltaNoise_vs_Temp}(b).
  It is possible to measure the $\delta\varepsilon$ noise up to somewhat higher temperatures and frequencies than the $\delta\Delta$ noise.
  Note that the $\Pi_\lambda$ values depend on the integration limits although the choice of $F_\lambda$ does not make any significant difference in the trends.
  (The double data points for the 165- and 180-mK $\delta\Delta$ noise were measured with different pulse spacings $\tau_0$.) 
  We estimated the device temperature from switching-current measurements on the SQUID, which suggest saturation at dilution-refrigerator temperatures below about 65\,mK.
  The error bars are derived only from the fit error of the read-out visibility $a(\tau_0,T)$, included in the $p_\mathrm{sw}$-to-$\delta \lambda$ conversion factor $\eta_\lambda(\tau_0,T)$, see Fig.~\ref{fig:deviceDescription}(f).
  }
\end{figure}

Finally, we measured the temperature dependencies of the two types of noise.
Figure~\ref{fig:tot_noise_power} shows the integrated noise powers $\Pi_\lambda$ vs.\@ temperature $T$ in the $65-200$-mK range, where our read-out visibility is sufficient.
We observe in essence temperature independence for both noises.
For $\delta\varepsilon$ (flux) noise, this is consistent with previous observations in SQUIDs~\cite{Wellstood-APL-87,Sendelbach-PRL-09};
we discuss the $\delta\Delta$ noise below.

\section{Discussion}

In order to analyze the $\delta\Delta$ noise, we parameterize it as an effective, normalized critical-current noise, $i_\mathrm{c} = \delta I_\mathrm{c}/I_\mathrm{c}$, with $I_\mathrm{c}=0.4\,\mu$A, in a Josesphson junction with area $\mathcal{A} = (0.2\,\mu\mathrm{m})^2$.
Van Harlingen et al.~\cite{vanHarlingen-PRB-04} found a ``canonical" value for the $1/f$ $\delta I_\mathrm{c}$-noise power at 1\,Hz and 4.2\,K:
$A_{I_\mathrm{c}}^\mathrm{can} \approx 144\,(\mathrm{pA})^2 (I_\mathrm{c}/\mu\mathrm{A})^2/(\mathcal{A}/\mu\mathrm{m}^2)$
in several SQUIDs and qubits of various sizes, made of different materials.
The authors hypothesized a quadratic temperature dependence, consistent with certain plausible models for the noise sources below 100\,mK, while noting that other models suggest a linear dependence.
The bilateral normalized noise-PSD then becomes $S_{i_\mathrm{c}}^\mathrm{can}(f) =
  A_{I_\mathrm{c}}^\mathrm{can} I_\mathrm{c}^{-2} \, (T/4.2\,\mathrm{K})^2 / |f|$,
which, for $T=65$\,mK, is considerably lower (almost 20 times) than our measured value.
On the other hand, Eroms et al.~\cite{Eroms-APL-06} measured resistance fluctuations in aluminum tunnel junctions:
they found about 100 times lower noise power at 4.2\,K, a linear temperature dependence, and saturation below 0.8\,K, i.e.,
$S_{i_\mathrm{c}}(f) =  (1/100) \times A_{I_\mathrm{c}}^\mathrm{can} I_\mathrm{c}^{-2} \, (T/4.2\,\mathrm{K}) / |f|$.
With $T=0.8$\,K, this gives a value about 2.5 times lower than what we observe.
We also note that recently, contrary to these findings, Paik et al.~\cite{Paik-PRL-11} reported no evidence for $1/f$ $i_\mathrm{c}$ noise in a Josephson junction.

An alternative source of $\delta\Delta$ noise is the fluctuating offset charges, $\delta Q$, known to exhibit $1/f$ noise~\cite{Bladh-PhysE-03,Astafiev-PRL-04,Zimmerman-PRB-09}; these charges effectively supply a gate voltage to each island.
The charge-noise power typically observed in single-electron tunneling (SET) devices is proportional to temperature~\cite{Gustafsson-arxiv-12} (although quadratic dependencies have also been observed~\cite{Astafiev-PRL-04}) and saturates below about 200\,mK, due to self heating of the SET, at a ``canonical" value $A_Q$ of about $(1 \!\sim\! 10\,\mathrm{m}e)^2$ at 1\,Hz.
We estimate our qubit's maximum sensitivity to charge fluctuations, $\kappa_{\Delta,Q} \equiv \partial\Delta/\partial Q$, to be in the range $0.1 \!\sim\! 1\,\mathrm{MHz}/e$.
We can then parameterize the $\delta\Delta$ noise as charge noise and estimate the dephasing time
$T_\varphi^* = (2\pi\,\kappa_{\Delta,Q} D_\Delta)^{-1} A_Q^{-1/2} (\ln[t_\mathrm{acq}/2\,\tau_0])^{-1/2} \approx 4 \sim 400\,\mu$s.
%
%
The lower end of this range is not far from our observed value.
%
%
Moreover, the tunneling of charged quasiparticles between the small islands constituting our device may displace offset charges and contribute to dephasing at $\varepsilon=0$.


In conclusion, our spectroscopy of both $\delta\varepsilon$ noise (flux noise) and $\delta\Delta$ noise (effective critical-current or charge noise), facilitated by single-shot measurements and thorough data analysis, shows that the very same $1/f^\alpha$ dependencies, measured at substantially higher frequencies, extend down to millihertz frequencies.
This apparently indicates that the same noise mechanisms are active and dominant over some ten orders of magnitude or more for $\delta\varepsilon$ noise and at least eight orders of magnitude for $\delta\Delta$ noise.
The $\delta\varepsilon$ noise may extend, with roughly constant slope (on a  logarithmic scale), up to the qubit's transition frequency at several gigahertz~\cite{Bylander-NPhys-2011}: there, this noise is nearly transverse to the flux qubit's energy eigenbasis, and would therefore also contribute to energy relaxation.
The small, if not negligible, $\delta\varepsilon - \delta\Delta$-noise correlations (over $7\times10^{-4}-2\times10^{-1}$\,Hz) show that the noises are due to distinct underlying mechanisms.
Moreover, both noises are temperature independent in the $65-200$-mK range, which suggests that the microscopic mechanisms are dominated by even lower energy scales than that.
This is useful information for the development of noise models. It also calls for further studies of the reproducibility of the device properties, and, in particular, of the $\delta\Delta$ noise, as it limits the coherence time in superconducting flux and transmon qubits.


\section*{Appendix}

\subsection{Spectral density and the statistical noise floor}

Here we describe how we calculate the noise-power spectral density (PSD) from the noisy time series, and eliminate the statistical white-noise floor due to sampling.

\subsubsection{PSD}

The fluctuations of our qubit's transition frequency constitute a zero-mean, wide-sense stationary process $\delta\nu(t)$;
at our chosen working point, $\delta\nu(t) = \nu(t) - 1/4\tau_0$.
We seek its bilateral noise-PSD (in units of rad/s, i.e., we use the angular-frequency correlator),
\begin{equation} \label{eq:psd_continuous}
    S(f) = (2\pi)^2 \, \lim_{T\to\infty}
    \frac{ \left\langle | V(f) |^2 \right\rangle }{T} .
\end{equation}
Our measurements' raw data, however, consists of a binary time series $\{z_n\}$ with elements of expectation value
$y_n = \langle z_n \rangle = p_0 + a_0 \sin(2\pi \,\delta\nu(n\Delta t) \,\tau_0)$,
where $\Delta t$ is the time step.
The statistical properties of $\{z_n\}$ represent those of the underlying process $\delta\nu(t)$, up to a conversion factor and a correction factor (explained in the next section).
We can therefore take the discrete Fourier transform
$\{Z_k\} = \mathcal{F}[\{z_n\}]$,
identify $Z(f_k) = Z_k \times \Delta t$ for $f_k=k/N\Delta t$, and compute the discrete, bilateral noise-PSD over the frequency range from $1/t_\mathrm{tot} = 1/N\Delta t=10$\,mHz to $1/2t_\mathrm{acq} = 1/2\Delta t=250$\,Hz,
\begin{equation} \label{eq:psd}
    S_{k=0} = (2\pi)^2 \, \frac{1}{2}
    \frac{Z_0^2 \, (\Delta t)^2}{N \Delta t} ,
    \qquad
    S_{k\neq 0} = (2\pi)^2 \, \frac{Z_k^2 \,(\Delta t)^2}{N \Delta t} ,
\end{equation}
$$
    \mathrm{for} \qquad
    k=0,\ldots,(N-1)/2
    \qquad \mathrm{with} \qquad
    f_k = \frac{k}{N\Delta t} .
$$
We then take the statistical average of $M$ different PSDs obtained from different time series measured in succession,
\begin{equation} \label{eq:Sxx_stat_avg}
    \left\langle S_k \right\rangle_\mathrm{stat} = \frac{1}{M} \sum_{m=1}^M S_k^{(m)} ,
\end{equation}
and finally smooth the result with a sliding average in the frequency domain.

With this method (Eqs.~\ref{eq:psd}--\ref{eq:Sxx_stat_avg}), each element $z_n$ is the result of a single-shot measurement;
the sampling time step $t_\mathrm{acq}$ is the same as the pulse-sequence repetition time $\Delta t$.
This sets it apart from the standard approach of first taking the ensemble average of typically $N_\mathrm{G}=1,000$ samples in the time domain, before calculating the PSD of the resulting $N/N_\mathrm{G}$ sampled points.
The acquisition time is then $t_\mathrm{acq} = N_\mathrm{G}\Delta t$, and the upper cut-off frequency becomes only $1/2t_\mathrm{acq}\approx 0.25$\,Hz.

\subsubsection{White-noise floor}

The PSD of the single-shot time sequence suffers from statistical sampling noise because each time step $n$ constitutes a Bernoulli trial ($b_n$):
the read-out SQUID switches ($b_n = 1$) with probability $p$ and does not switch ($b_n = 0$) with probability $1-p$.
This statistical noise has a white spectrum;
it dominates possible white background noise from other sources, and dominates also the $1/f$ noise at high frequencies.
To estimate it, we can treat the stochastic variable $b_n$ as independent and identically distributed (i.i.d.\@) with ensemble-averaged mean $\langle b\rangle=p$ and variance $\sigma_b^2 \equiv \langle (\Delta b)^2 \rangle \equiv
\langle b^2\rangle - \langle b \rangle^2 = p\,(1-p)$.
Sampling at a fixed rate $1/\Delta t$, the white-noise floor of the bilateral PSD becomes
\begin{equation} \label{eq:modified_Carson_Bernoulli}
  S_\mathrm{n}(f_k) = (2\pi)^2
        \left( \sigma_b^2 + \langle b\rangle^2 \delta_{k,0} \right) \, \Delta t .
\end{equation}
Here we use Kronecker's delta $\delta_{k,0}$ in the discrete PSD.

The same expression is valid for the PSD of the ensemble-averaged time series, the constituent elements of which have a binomially distributed switching probability ($c$ for ``counts") averaged over a gate time $t_\mathrm{acq} = N_G\Delta t$\,:
we obtain Eq.~(\ref{eq:modified_Carson_Bernoulli}) after substituting $\langle c\rangle = p$ and $\sigma_c^2 = p\,(1-p)/N_G$ for $\langle b\rangle$ and $\sigma_b^2$, respectively.

Equation~(\ref{eq:modified_Carson_Bernoulli}) is, in fact, a modification of Carson's theorem, which is valid for temporally random pulse arrivals.
There, one considers a random pulse train
$w(t) = \sum_{l=1}^L b_l \, g(t-t_l)$, in which $g(t)$ is the pulse envelope, the stochastic variable $b_l$ is the (continuous) pulse height, and the stochastic variable $t_l$ is the pulse-arrival time.
The Fourier transform of $w(t)$ is
$W(f) = G(f) \sum_{l=1}^L b_l \exp(-i2\pi f t_l)$,
where $G(f) = \mathcal{F}[g(t)]$.
Carson's theorem is then
\begin{equation} \label{eq:standardCarson}
S_\mathrm{n}^\mathrm{Carson}(f_k) = (2\pi)^2 \left( \langle 1/\Delta t\rangle \langle b^2 \rangle |G(f_k)|^2 + \langle b \rangle^2 \delta(f_k) \right) .
\end{equation}
In our case, the pulse height $b_l$ is binary and the pulse-arrival rate is fixed at $1/\Delta t$; we can therefore write (with Kronecker's delta)
$G = \mathcal{F}[\delta_{m,0}] = 1\times\Delta t$.
We just have to replace the mean-square $\langle b^2 \rangle$ by the variance $\sigma_b^2$ and set $\delta(f_k) = \delta_{k,0}\Delta t$ to obtain Eq.~(\ref{eq:modified_Carson_Bernoulli}).

Parenthetically, one can also derive Eq.~(\ref{eq:modified_Carson_Bernoulli}) by using the Wiener--Khintchine theorem.
The autocorrelation function is
\begin{equation} \label{eq:autocorr-fcn}
R_{bb}^{(m)} = (1/N)\times \sum_{n=0}^{N-1} b_n b_{n-m} = \langle b \rangle^2 + \sigma_b^2\,\delta_{m,0} ,
\end{equation}
and $\mathcal{F}[1]=\delta_{k,0}\times\Delta t$, so that
\begin{equation} \label{eq:standardCarson_2}
S_\mathrm{n}(f_k) = (2\pi)^2 \mathcal{F}
\left[ R_{bb}^{(m)} \right] = \mathrm{Eq.}~(\ref{eq:modified_Carson_Bernoulli}).
\end{equation}

We find that the PSD resulting from a simulation of Bernoulli- and binomially distributed noise agrees well with the measured data and with Eq.~(\ref{eq:modified_Carson_Bernoulli}):
we therefore conclude that our experimental noise floor is due to the statistical sampling.

If the data consisted of a train of pulses of finite length in time, the PSD would have a roll-off near the Nyquist frequency $1/2\Delta t$.
For example, the Fourier transform of a boxcar (square) pulse of length $\Delta t$ is the function $\Delta t \, \mathrm{sinc}\left(\pi f \Delta t \right)$.
In our case, after conversion of the SQUID's response (the presence or absence of a voltage pulse) to binary form, our data can be seen as represented by a train of delta-functions, and their Fourier transform is frequency independent, i.e., our white noise floor has no roll-off.

\subsubsection{Cross-PSD: white-noise elimination}

In order to eliminate the white-noise floor, at the expense of a halved Nyquist frequency, we calculate the discrete cross-PSD of interleaved time series, i.e., by setting $z'_n=z_{2n-1}$ and $z''_n=z_{2n}$ (with $n=1,\ldots,N/2$) and computing the cross spectrum of $z'_n$ with $z''_n$.
(We again assume that the stochastic switching process is uncorrelated from sample to sample, at frequency $1/\Delta t$.)
The resulting PSD is
\begin{equation} \label{eq:cross_psd}
    S^\mathrm{cross}_{k=0} = (2\pi)^2 \, \frac{1}{2}\,
    \frac{Z'_k\,(Z''_k)^{^*}}{N/2\Delta t} ,
    \qquad
    S^\mathrm{cross}_{k\neq0} = (2\pi)^2 \,
    \frac{Z'_k\,(Z''_k)^{^*}}{N/2\Delta t} ,
\end{equation}
where $k = 0,\ldots,N/4$ and $f_k = k/N\Delta t$.
These expressions reproduce the noise spectrum, with the use of correction factors, as explained in the following section (Eqs.~\ref{eq:crossPSDexpectation}--\ref{eq:finalCorrFactor}).

Compared to the previous section, we have eliminated the white noise by circumventing the zero-delay autocorrelation term $R_{bb}^{(0)}$ in Eq.~(\ref{eq:autocorr-fcn}), and are only left with the delta-function component,
\begin{equation} \label{eq:cross_psd_floor}
    S^\mathrm{cross}_\mathrm{n}(f_k) = (2\pi)^2 \, \langle b \rangle^2 \, \delta_{k,0} \Delta t .
\end{equation}

In the same way as in Eqs.~(\ref{eq:autocorr-fcn}--\ref{eq:standardCarson_2}), this can be derived by applying the Wiener--Khintchine theorem to the cross correlation function, which this time simply gives $R_{b'b''}^{(m)} = \langle b \rangle^2$ (the subscript $b'b''$ indicates two interleaved sub-series obtained from the original series $b_n$), and therefore
\begin{equation} \label{eq:WK-2}
    \quad S^\mathrm{cross}_\mathrm{n}(f_k) = (2\pi)^2 \,
    \mathcal{F} \left[ R_{bb'}^{(m)} \right] =  \mathrm{Eq.}~(\ref{eq:cross_psd_floor}) .
\end{equation}

When calculating the PSD, we take the statistical average of $S_k^\mathrm{cross}$, keeping the averaging coherent throughout (i.e.\@ retaining $S_k^\mathrm{cross}$ as a complex quantity), and, just as for $S_k$, smoothen it with a sliding average before plotting its magnitude
$\left|\langle\langle S_k^\mathrm{cross} \rangle_\mathrm{stat}\rangle_\mathrm{fq}\right|$.
%

The result (\ref{eq:cross_psd}) is equivalent to the explicit subtraction of the incoherent noise $S_\mathrm{n}$ from $S_k$ (Eqs.~\ref{eq:psd}, \ref{eq:modified_Carson_Bernoulli}), with the advantage, however, of drastically reduced uncertainty, in particular at high frequencies where the ($1/f$-noise) signal is much smaller than the white noise.
This method is appropriate for the analysis of, e.g., $1/f$-type noise.
However, it is not applicable in a predominantly white-noise environment: then, the noise under study would be eliminated along with the statistical white noise.

\subsection{Correction factors: Quasi-static noise and the non-linear transfer function}

In this section, we treat the effects on the PSD caused by quasi-static noise, and by the sine nonlinearity in the conversion from the measured switching events to the variations of the qubit's transition frequency.
%

\subsubsection{Decay of the Ramsey fringe -- quasi-static noise}

Noise in the effective longitudinal field coupled to the qubit results in decoherence of the quantum superposition.
We denote a fluctuation as ``quasi-static" or ``incoherent" noise, when it can be considered as static during each free-induction period, but varying over the longer time span between experimental realizations.
Dephasing results from such uncorrelated fluctuations of the Larmor frequency $\nu_{01}$, and therefore of the accrued phase of the superposition state, $\varphi(\tau) = 2\pi \int_0^\tau \mathrm{d}t \, \nu_{01}(t)$.
It leads to decay of the Ramsey free-induction signal, as each measured point is the incoherent average of many experimental realizations.
We describe this fluctuation by a standard deviation,
\begin{equation} \label{eq:sigma}
\sigma^2 = 2 \int_{1/t_\mathrm{acq}}^{1/\tau} \mathrm{d}f\,S(f) .
\end{equation}
The higher integration limit is here the inverse of the free-induction time, $1/\tau\approx 0.1-100$\,MHz;
fluctuations at even higher frequencies are effectively canceling out.
The lower limit is given by the total acquisition time $t_\mathrm{acq}$ used to infer the qubit's population at each fixed free-induction time span $\tau$.
Typically averaging over $N_\mathrm{avg}=5,000$ measurements with a repetition time $\Delta t = 2\,$ms, we obtain $t_\mathrm{acq} = N_\mathrm{avg}\Delta t = 10$\,s.
(If instead the measurements were done in the opposite order, stepping over $\tau$ in the inner loop, with $N_\mathrm{pts}\approx 100$ steps, and averaging over $N_\mathrm{avg}$ in the outer loop, the total acquisition time would be $N_\mathrm{pts} N_\mathrm{avg} \Delta t = 1,000$\,s, and the lower cut-off frequency would be correspondingly lower.)

Ensemble averaging over all realizations of $\delta\varphi(\tau)$, and assuming Gaussian fluctuations resulting from numerous fluctuators, we obtain the dephasing envelope
\begin{equation} \label{eq:dephasingIntegral}
h(\tau) = \langle \exp(i\delta\varphi(\tau)) \rangle = \exp(-\langle(\delta\varphi)^2\rangle/2) =
\end{equation}
$$
= \exp\left( - \tau^2 \int_{f_\mathrm{low}}^{f_\mathrm{high}} \mathrm{d}f \, S(f) \, \mathrm{sinc}^2(\pi f \tau) \right) \approx
$$
$$
\approx \exp\left(- \sigma^2\,\tau^2/2\right),
$$
where the sinc-squared function is due to the square time window of the Ramsey pulse sequence, and we can approximate it by unity for $f<1/2\tau$.

As an illustration, we now evaluate $h(\tau)$ for the two cases of $1/f$ noise and white noise.
For $1/f$ noise, $S(f) = A/f$, Eq.~(\ref{eq:sigma}) becomes
$\sigma^2 = 2\, A \log\left[(1/2\tau)/(1/t_\mathrm{acq})\right]$.
The weak, logarithmic sensitivity to the cut-off frequencies effectively allows us to treat it as a time-independent constant, $\sigma^2 \approx 2\,A \,C$, giving Gaussian decay $p_{1/f}(\tau) = \exp(-\sigma^2\,\tau^2/2)$.
For white noise, $S(f) = S_\mathrm{w}$, on the other hand, the integral is linearly sensitive to the upper cut-off frequency, so that
$\sigma^2 = S_\mathrm{w}/\tau$,
yielding an exponential decay
$p_\mathrm{w}(\tau) = \exp(- S_\mathrm{w}\tau/2)$.
Here the exponent is proportional to time; we can therefore identify $1/T_\varphi^* = S_\mathrm{w}/2$ as the dephasing rate.

\subsubsection{Repeated fixed-time free-induction}

The previous section described how quasi-static noise determines the dephasing of the Ramsey-fringe.
Now we turn to its effect on Ramsey interference with a fixed free-induction time $\tau_0$, repeated numerous times.

With our single-shot measurements, each element of the binary time series $\{z_n\}$ is a Bernoulli random variable $z_n$ with expectation value given by the switching probability $p_\mathrm{sw}$, which we now denote as
\begin{equation} \label{eq:Psw}
y_n = p_0 + a_0 \sin x_n .
\end{equation}
This function has a non-linear dependence on $x_n = 2\pi \delta\nu_n \tau_0$, the phase accrued during $\tau_0$, where $\delta\nu_n$ is the average fluctuation of the transition frequency at time step $n$.
This phase $x_n$, in turn, has noise contributions from two distinct frequency intervals, ``1" and ``2."

We denote as ``interval 1" the frequencies which we can resolve by taking the Fourier transform of the series $\{z_n\}$, of total length $N\Delta t$ and step size $\Delta t$, i.e.\@ from $1/t_\mathrm{tot} = 1/N\Delta t\approx 10^{-2}$\,Hz to $1/2t_\mathrm{acq} = 1/2\Delta t\approx 250\,$Hz (or with the interleaving method up to $1/4\Delta t\approx 125$\,Hz).
The noise within this interval has zero mean and variance $\sigma_1^2$ (Eq.~\ref{eq:sigma}).

\begin{figure}     
\centering
\vspace{-4mm}
\includegraphics[width=7cm]{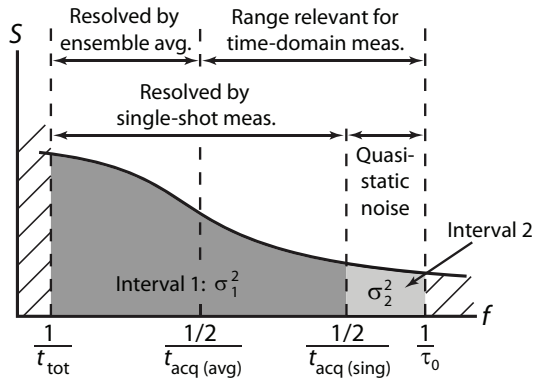}
  \caption{\label{fig:PSD_Sketch}
  Sketch of the PSD, indicating the frequency intervals resolved by the ensemble-averaging and single-shot sampling methods.
  Also indicated are the Gaussian, quasi-static noise and the variances $\sigma_{1,2}^2$ (Eq.~\ref{eq:sigma}).
  Here $t_\mathrm{tot}$ is the total length of the time trace (can be several minutes to hours); $t_\mathrm{acq(avg)}=N_\mathrm{G}\Delta t = 1\sim10$\,s is the acquisition time per measured point in time-domain experiments such as Ramsey and spin-echo decay;
  $t_\mathrm{acq(sing)}=\Delta t=2$\,ms is the repetition time (acquisition time of the single-shot samples);
  and $\tau_0\approx 1\,\mu$s is the pulse spacing.
}
\end{figure}

In addition, there is a contribution from the quasi-static noise in ``interval 2," which is the range from $1/2t_\mathrm{acq}$ to $1/\tau_0$;
see Fig.~\ref{fig:PSD_Sketch}.
This noise cannot be resolved, but acts in aggregate and leads to dephasing, e.g.\@ in a Ramsey-fringe experiment.
It has zero mean and variance $\sigma_2^2$ (Eq.~\ref{eq:sigma}).

Noise at even higher frequencies than $1/\tau_0$ averages out during free induction.

At each time step $n$, the element $x_n$ is subject to noise contributions from both intervals, and their variances add up to $\sigma_2 = \sigma_1^2 + \sigma_2^2$. 
We write $x_n = u_n + v_n$, where $u$ and $v$ refer to the noise originating in intervals 1 and 2, respectively.
Here $u_n$ has correlations between the different time steps $n$ due to the memory effect of the $1/f$ noise; on the other hand, $v_n$ is incoherent and can be taken as a Gaussian i.i.d.\@ random variable.

While it is impossible to unequivocally infer $x_n$ from the measured $z_n$ at each instance $n$, we can infer statistical properties of $\{x_n\}$, such as its correlations and spectral density, up to the frequency $1/2t_\mathrm{acq} = 1/2\Delta t$, which can approach $1/\tau_0$.
%
This is advantageous compared to the ensemble-averaging method, which has a longer acquisition time $t_\mathrm{acq}=N_\mathrm{G}\Delta t$.
%
%

We can write the $m\neq n$ autocovariance function for $\Delta z_n = z_n - \langle z_n \rangle$ as
\begin{equation} \label{eq:crossPSDexpectation}
\langle \Delta z_m\, \Delta z_n \rangle =
\langle \Delta y_m\, \Delta y_n \rangle =
a_0^2 \langle \sin x_m \sin x_n \rangle \simeq
\end{equation}
$$
\simeq a_0^2 \langle x_m\, x_n \rangle =
a_0^2 \langle u_m\,u_n \rangle .
$$
The first equality holds because the Bernoulli trials are independent, and the last equality is the consequence of  $v_n$ being i.i.d., which implies
$\langle u_m\,v_n \rangle = \langle v_m\,v_n \rangle = 0$.
%
%
%
The third step is an equality only when $|x_n|\ll 1$;
when $\sigma_2$ is large, e.g.\@ at higher temperatures, or when we use a larger free-induction time $\tau_0$ to decrease the statistical noise level, the variation of $x_n$ can be large, and then this is not a good approximation.
%
%
Instead of approximating, however, we can compensate the result for the sine nonlinearity.
Expanding the correlator $\langle \Delta y_m \Delta y_n \rangle$, we obtain
\begin{equation} \label{eq:sinesine}
\langle \sin x_m \sin x_n \rangle =
\langle \sin (u_m+v_m) \sin (u_n+v_n) \rangle =
\end{equation}
$$
= \langle (\sin u_m \cos v_m + \cos u_m \sin v_m) (m\to n)\rangle .
$$
Since sine is an odd function and $v_n$ is a zero-mean, Gaussian i.i.d.\@ variable, $\langle \sin v_{m,n} \rangle = 0$, and (\ref{eq:sinesine}) becomes
\begin{equation} \label{eq:sinesineSimplified}
\langle \sin u_m \cos v_m \sin u_n \cos v_n \rangle =
\end{equation}
$$
= \langle \cos v_m \rangle
\langle \cos v_n \rangle
\langle \sin u_m \sin u_n \rangle .
$$
%
%

The cosine factors depend on noise in interval 2, i.e., above the sampling frequency.
This is similar to dephasing due to quasi-static noise, which acts uniformly on all the samples in time (incoherent averaging over a distribution of the noise), and leads to Gaussian decay functions
\begin{equation} \label{eq:incoherentCorrFactor}
\langle \cos v_{m,n} \rangle = \exp(-\sigma_2^2\tau_0^2/2) .
\end{equation}
%

For the sine factor, the noise is from interval 1, i.e., it is resolved by our sampling, and therefore is not uniform.
The process is a combination of ensemble-averaged incoherent noise and a frequency-dependent filtering due to the $(m-n)\Delta t$ time difference in the correlator.
Evaluating this factor, we obtain Gaussian damping of a hyperbolic-sine function of the correlator,
\begin{equation} \label{eq:sinesine2}
\langle \sin u_m \sin u_n \rangle =
\end{equation}
$$
=  \int\!\!\!\!\int\!\mathrm{d}u_m \mathrm{d}u_n \, \sin u_m \sin u_n \, N(0, \mathbf{\sigma}) =
$$
$$
= \exp(-\sigma_1^2\tau_0^2) \sinh(\langle u_m u_n \rangle) ,
$$
where the integral is taken over a two-dimensional normal distribution $N(0,\sigma)$ with zero mean and correlation matrix
$\mathbf{\sigma} = \{ \sigma_m, \sigma_n, \sigma_{mn} \}$.
(The distribution widths are equal, $\sigma_m=\sigma_n$, and $\sigma_{mn} = \langle u_m u_n \rangle$ is the correlation function.)

The correlator~(\ref{eq:crossPSDexpectation}) finally becomes
\begin{equation}\label{eq:CorrFactor}
\langle \Delta z_m \Delta z_n \rangle =
\end{equation}
$$
= a_0^2 \exp(-\sigma_1^2\tau_0^2) \exp(-\sigma_2^2\tau_0^2)
\sinh( \langle u_m u_n \rangle ) .
$$
Note that no approximation has been made so far (cf.\@ Eq.~\ref{eq:autocorr-fcn}).
If the noise correlation due to $1/f$-type noise is small, as in our case where $\exp(\sigma_1^2 \tau_0^2) < 10$, we can neglect the frequency-dependent filtering effect and approximate
$\sinh( \langle u_m u_n \rangle ) \approx \langle u_m u_n \rangle$.

Now remains only the determination of the correction factors, which we know from the calibration measurement,
$\exp\left[(\sigma_1^2+\sigma_2^2)\tau_0^2\right] = (a_0/a(\tau_0))^2$, where we identify $a(\tau_0)/a_0 = h(\tau_0)$ (Eq.~\ref{eq:dephasingIntegral}), so that, finally,
\begin{equation} \label{eq:finalCorrFactor}
\langle u_m u_n \rangle \approx
\langle \Delta z_m \Delta z_n \rangle / \left(a(\tau_0)\right)^2.
\end{equation}
We note that it resembles the signal damping due to dephasing in a Ramsey fringe.
The actual numbers used in our analysis of the data in Figs.~2 and 4 are presented in Tables~\ref{tab:eps}--\ref{tab:Delta2}.
    \begin{table}[!h]
        \caption{$\delta\varepsilon$ noise ($\varepsilon = 640$\,MHz).
        Data in Figs.~2 and 4.}
        \vspace{2mm}
        \begin{tabular}{c|c|c}
            Temp. (mK) \, & \, $\tau_0$ (ns) \, & \, $\exp(\sigma^2\tau_0^2)$ \\
            \hline
            \,\,\,65 & 50 & 1.5    \\
            120 & 50 & 1.6    \\
            165 & 50 & 1.6    \\
            210 & 50 & 1.8
        \end{tabular}
    \label{tab:eps}

    \end{table}
    \begin{table}[!h]
    \caption{$\delta\Delta$ noise ($\varepsilon = 0$).
    Data in Figs.~2 and 4.}
        \vspace{2mm}
        \label{tab:Delta1}
        \begin{tabular}{c|c|c}
            Temp. (mK) \, & \, $\tau_0$ (ns) \, & \, $\exp(\sigma^2\tau_0^2)$ \\
            \hline
            \,\,\,65 & 300 & 1.2 \\
            120 & 300 & 1.3 \\
            165 & 1,200 & 7.4 \\
            180 & 1,000 & 5.6
        \end{tabular}
    \end{table}
    \begin{table}[!h]
    \caption{$\delta\Delta$ noise ($\varepsilon = 0$).
    Data in Fig.~4 (but not in Fig.~2).}
        \vspace{2mm}
    \label{tab:Delta2}
        \begin{tabular}{c|c|c}
            Temp. (mK) \, & \, $\tau_0$ (ns) \, & \, $\exp(\sigma^2\tau_0^2)$ \\
            \hline
             165 & 300 & 1.3 \\
             180 & 300 & 1.6
        \end{tabular}
    \end{table}

\newpage

\subsection{Data smoothing and reproducibility of the PSD}

The following Figs.~\ref{fig:DeltaNoise_repeat_165mK}--\ref{fig:DeltaNoise_repeat_180mK} show the reproducibility of our results, with sets of data taken on different days.
Figures~\ref{fig:DeltaNoise_vs_Q}--\ref{fig:EpsNoise_vs_Q} show that our PSD's power laws are independent of the choice of smoothing windows.

\begin{figure}[!h]     
\centering
\includegraphics[width=8cm]{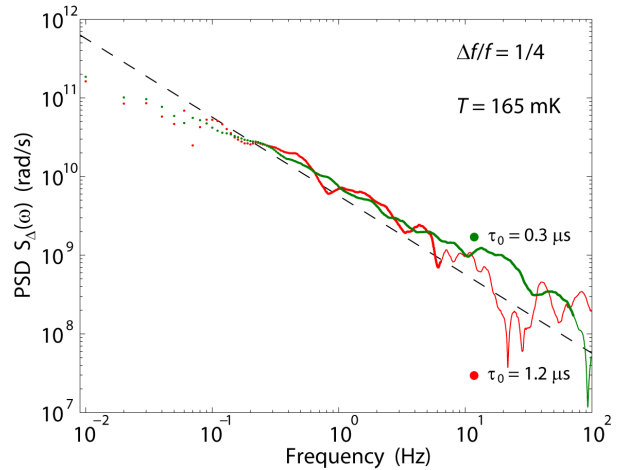}
  \caption{\label{fig:DeltaNoise_repeat_165mK}
  (color online) $\delta\Delta$ noise at 165\,mK with two different pulse spacings $\tau_0$, showing reproducibility of the noise-PSD; cf.\@ Tables~\ref{tab:Delta1}--\ref{tab:Delta2}.
}
\end{figure}

\begin{figure}[!h]     
\centering
\includegraphics[width=8cm]{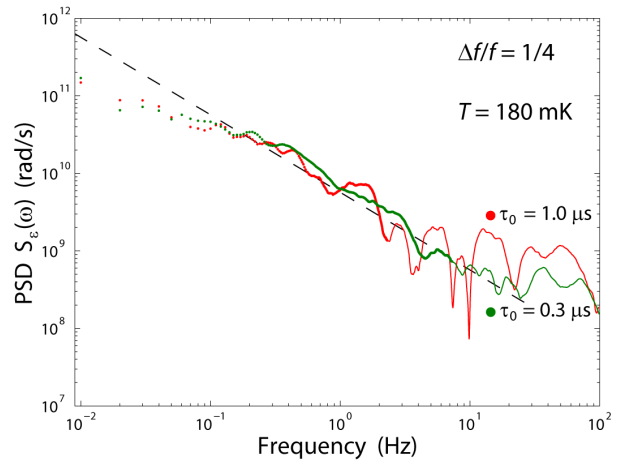}
  \caption{\label{fig:DeltaNoise_repeat_180mK}
  (color online) $\delta\Delta$ noise at 180\,mK, otherwise like Fig.~\ref{fig:DeltaNoise_repeat_165mK}.
}
\end{figure}

\begin{figure*}[h]     
\centering
\includegraphics[width=14cm]{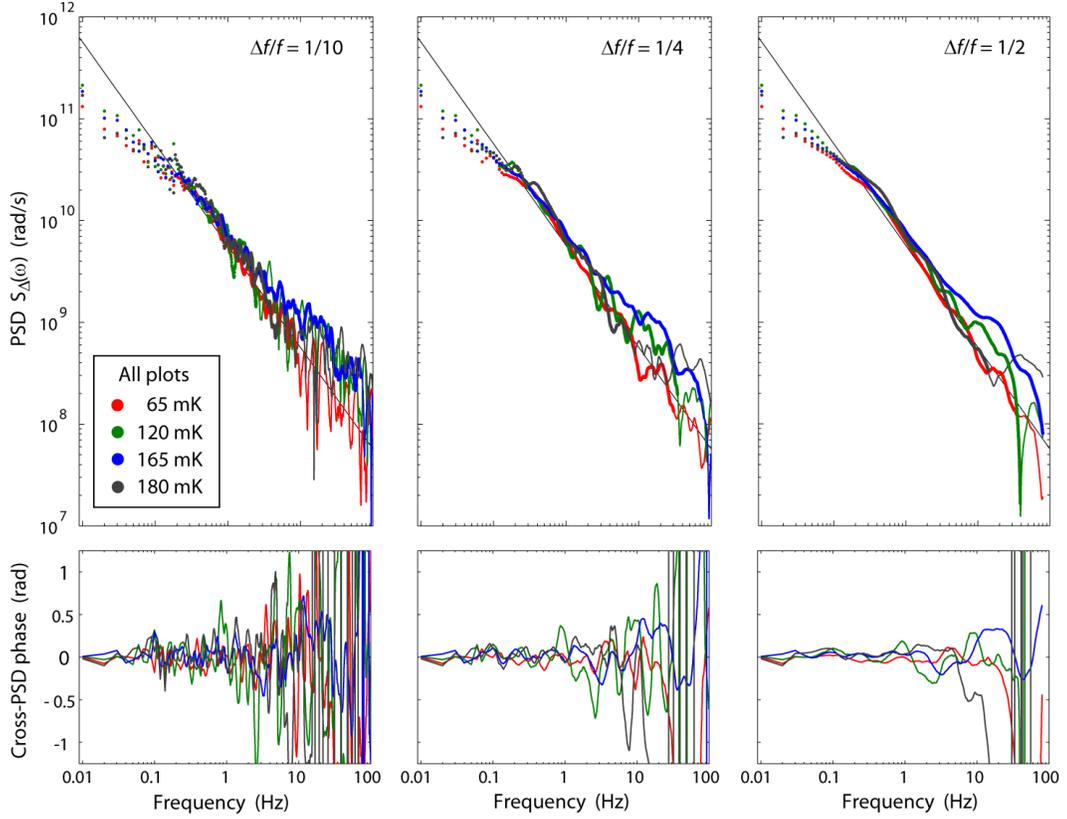}
  \caption{\label{fig:DeltaNoise_vs_Q}
  (color online) $\delta\Delta$ noise with different smoothing windows $\Delta f/f$.
  We choose the upper cut-off frequency $f_\mathrm{c}$ for Fig.~2 as the lowest frequency for which the phase of the cross-PSD deviates from zero by more than 1 rad.
  In that figure we use the smoothing window $\Delta f/f=1/4$. The spectrum displays no significant difference depending on $\Delta f/f$, and the structure can be attributed to insufficient averaging.
  The phase deviation is, also, due to insufficient averaging, and becomes larger for increasing temperature, for a fixed pulse separation $\tau_0$.
}
\end{figure*}

\begin{figure*}[h]     
\centering
\includegraphics[width=14cm]{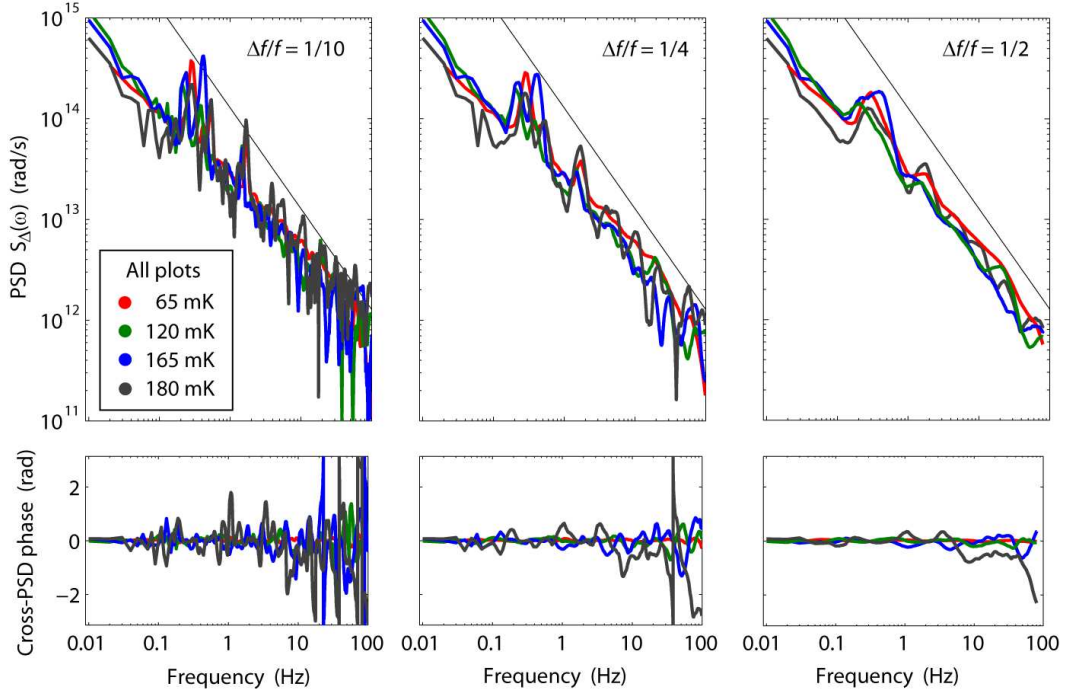}
  \caption{\label{fig:EpsNoise_vs_Q}
  (color online) $\delta\varepsilon$ noise with different smoothing windows $\Delta f/f$, cf.\@ Fig.~\ref{fig:DeltaNoise_vs_Q}.
}
\end{figure*}

\newpage

\section*{Acknowledgements}

We acknowledge discussions with  G.\@ Chen, L.\@ DiCarlo, M.\@ Gustafsson, X.\@ Jin, and L.\@ Wang.
We thank the LTSE team at MIT Lincoln Laboratory for technical assistance.
This work was sponsored in part by the U.S.\@ Government, the Laboratory for Physical Sciences, the U.S.\@ Army Research Office (W911NF-12-1-0036), the National Science Foundation (PHY-1005373), and the Funding Program for World-Leading Innovative R\&D on Science and Technology (FIRST), NICT Commissioned Research, MEXT kakenhi ``Quantum Cybernetics."
Opinions, interpretations, conclusions and recommendations are those of the author(s) and are not necessarily endorsed by the U.S.\@ Government.

\newpage

%
%
%
%
%
%
%
%
%
%
%
%
%
%






\end{document}